\documentclass[11pt]{article}
\usepackage[utf8]{inputenc}
\usepackage{color}
\usepackage{comment}
\usepackage{amsthm}
\usepackage{amssymb}
\usepackage{amsmath}
\usepackage{float}
\usepackage{bm}
\usepackage{subcaption}

\usepackage{graphicx}
\graphicspath{{figs/}}

\begin{document}
\title{
Numerical Investigation of Single-Core to Split-Core Transitions in Nematic Liquid Crystals\thanks{D.S. and P.L. acknowledge partial support from the National Science Foundation under Grant No. DMS-2318053. The authors are deeply grateful to Prof. M. C. Calderer (University of Minnesota) for her insightful guidance and constructive discussions that significantly contributed to this study.}}

\author{Daniel Siebel-Cortopassi \footnotemark[1]
\and Pei Liu \footnotemark[1]}

\footnotetext[1]{Department of Mathematics and Systems Engineering, Florida Institute of Technology, Melbourne, FL, 32901. 
Corresponding author: pliu@fit.edu.}
\date{}
\maketitle

\begin{abstract}
We analyze single-core and split-core defect structures in nematic liquid crystals within the Landau–de Gennes framework by studying minimizers of the associated energy functional. A bifurcation occurs at a critical temperature threshold, below which both split-core and single-core configurations are solutions to the Euler-Lagrange equation, with the split-core defect possessing lower energy. Above the threshold, the split-core configuration vanishes, leaving the single-core defect as the only stable solution. We analyze the dependence of such temperature threshold on the domain size and characterize the nature of the transition between the two defect types. We carry out a quantitative study of defect core sizes as functions of temperature and domain size for both single and split core defects. 
\end{abstract}

\section{Introduction}

Liquid crystals represent a mesophase of matter that combines fluidity with long-range orientational order, leading to rich and complex behavior not found in conventional phases. In particular, their partial ordering gives rise to a variety of topological defects\--regions where the molecular alignment becomes singular or indeterminate. Such defects play crucial roles to both their physical properties and technological applications. For example, topological defects in the liquid-crystalline packing of bacteriophage DNA reveal how confinement, elasticity, and molecular interactions govern genome organization, providing key insights into viral DNA structure and function \cite{leforestier2010bacteriophage,walker2020liquid,hiltner2021chromonic,liu2021ion,liu2022helical}. Similarly, the ability to manipulate defect configurations controls molecular alignment, stability, and the release kinetics of encapsulated drugs and can influence drug crystallization pathways \cite{teerakapibal2018organic,pisano_liquid_2019,shete_liquid_2021,yu2022structures}. In engineered materials including flat-screen displays, lasers and stretchable electronics, precise control over defect structures is essential to achieving optimal image performance \cite{morris_removing_2009,mysliwiec_liquid_2021,castles_stretchable_2014,lee_introduction_2020,lavrentovich2020design,lavrentovich2020ferroelectric}. Optical phenomena arising from the presence of topological defects, including blue phase behavior, Schlieren textures, and fingerprint textures among others, remain central to both the understanding and the control of liquid crystal behavior \cite{brinkman_defects_1982,gennes_physics_1993,oswaldNematicCholestericLiquid2005,dilisi_history_2019}. Depending on molecular structure, environment, and boundary conditions, liquid crystals can exhibit nematic, smectic, or cholesteric phases, each supporting multiple defect types. 

Among the earliest defect types to be experimentally observed were line disclinations \cite{friedelEtatsMesomorphesMatiere1922}, although their theoretical characterization emerged only later. Some of the earliest experimental works regarding point defects in liquid crystals were performed by Melzer and Nabarro in the 1970's \cite{melzerColsNoeudsNematic1977,melzerOpticalStudiesNematic1977}. Topological defects in liquid crystals are generally classified as one of these two types\--line defects (disclinations) and point defects, and arise in various combinations depending on boundary conditions and material geometry. Regarding point defects located away from the domain boundary, these only rarely \cite{chuangCoarseningDynamicsUniaxial1993} occur upon quench (rapid temperature decrease from the isotropic state into the nematic state). In liquid crystal colloids, with the liquid crystal being either the medium \cite{leal-calderonProgressUnderstandingEmulsion1999,starkPhysicsColloidalDispersions2001} or the particles suspended in another medium \cite{drzaicLiquidCrystalDispersions1995}, point defects can be observed throughout the colloid whenever the anchoring at the material interfaces is strong enough \cite{klemanTopologicalPointDefects2006}. When a material contains only two point defects separated by some distance, this configuration is called a ``split-core defect".
 
The equilibrium configurations of liquid crystals and their associated defect structures are most commonly described by two continuum models: the Oseen-Frank model \cite{oseen_theory_1933, frank_i_1958} and the Landau-de Gennes model \cite{degennesShortRangeOrder1971, gennes_physics_1993}. In Oseen-Frank theory, the liquid crystal is represented by a unit vector field $\mathbf{n}$, known as the director, which describes the local average molecular alignment. Owing to the head-to-tail symmetry of typical liquid crystal molecules, the director satisfies $\mathbf{n} = -\mathbf{n}$. This model captures coarse-scale behavior away from defects effectively, but it cannot represent line defects or biaxial order; in particular, line defects carry formally infinite energy within the classical Oseen-Frank framework. To overcome this limitation, Ericksen \cite{ericksenConservationLawsLiquid1961, ericksenHydrostaticTheoryLiquid1962} extended the model to allow a variable degree of orientation, enabling the representation of line defects while retaining the continuum description. In contrast, Landau-de Gennes theory can describe both uniaxial and biaxial states and remains well-defined in all defect cores. Where point defects and line disclinations in the Oseen-Frank model are singularities of the director field, they are not singularities \cite{gartland_scalings_2018,dassbach_computational_2017} in the corresponding tensor field used for the Landau-de Gennes model. However, some of the parameters related to the Landau-de Gennes energy can at times go out of range, yielding nonphysical configurations. Ball and Majumdar \cite{ball_nematic_2010} derive an energy functional which combines elements from mean-field Maier-Saupe energy and the Landau-de Gennes energy and results in physically realistic parameters for all temperature ranges, but this is beyond the scope of this paper. The Landau-de Gennes model characterizes the liquid crystal using a symmetric, traceless tensor $Q$, whose eigenvalues measure the degree of molecular ordering and whose eigenvectors indicate the preferred orientations. Let $\mu(\mathbf{p}, \mathbf{x})$ represent the probability measure that a molecule at position $\mathbf{x}$ points in the direction $\mathbf{p} \in \mathbb{S}^2$, and define $M := \int_{\mathbb{S}^2} \mathbf{p} \otimes \mathbf{p} \, d\mu(\mathbf{p},\mathbf{x})$, representing the second moment (or variance) of the molecular orientation, and $M_0 := \frac{1}{4\pi} \int_{\mathbb{S}^2} \mathbf{p} \otimes \mathbf{p} \, dA$, the second moment tensor of an isotropic orientation distribution; then $Q$ is defined as $M - M_0$. Compared with the Oseen-Frank and Ericksen theories, Landau-de Gennes provides a more complete and flexible framework for modeling localized defect behavior, biaxiality, and phase transitions, making it particularly suitable for studying the structure and energetics of complex defect cores.


In the literature, extensive efforts have focused on studying various types of defects in nematic liquid crystals confined to cylindrical or spherical domains with homeotropic (perpendicular) anchoring, as commonly found in liquid crystal droplets and colloidal systems. Depending on intrinsic material properties, temperature and domain geometry, three principal defect configurations have been identified: the single-core (hedgehog) defect, the split-core defect, and the Saturn ring defect \cite{gartland_structures_1995}.
The single-core defect consists of a single point singularity located at the center of the domain. The split-core defect \cite{mkaddem_fine_2000} is comprised of two distinct point defects connected by a short disclination line segment, forming a configuration that breaks radial symmetry. The Saturn ring defect, which consists of a closed disclination loop encircling the central axis of the domain, is a configuration that ``escapes" into the third dimension \cite{kralj_biaxial_1999}. In domains other than spherical and cylindrical, additional defect types \cite{quTransitionDefectPatterns2017} can be found.

Homeotropic boundary conditions in such domains enforce the presence of topological point defects \cite{penzenstadler_fine_1989}, making the analysis of their structure and transitions essential to understanding defect-related phenomena in liquid crystal systems. The split-core defect and the Saturn ring defect in the class of axially symmetric solutions have been studied numerically in \cite{hu2016disclination,mkaddem_fine_2000}. Under this constraint, these two defect types have recently been constructed in a rigorous manner \cite{dipasqualeToruslikeSolutionsLandaude2024,taiPatternFormationLandau2023,yuDisclinationsLimitingLandau2020}. In \cite{yuDisclinationsLimitingLandau2020}, they go on to investigate the relationship between two- and three-dimensional defects confined in a shell, and find that when the thickness of the shell is small, the solution is radial-invariant, and when the thickness of the shell increases, symmetry of the solution is broken. In the same work they also prove that within all axially symmetric maps in the unit sphere, the Landau-de Gennes energy functional $E_{LDG}$ has at least two distinct energy minimizers: the Saturn ring and the split core. \cite{gengUniformProfilePoint2022} recently rigorously demonstrated that when there is one point defect, outside of the defect core the Landau-de Gennes minimizer can be well approximated by the Oseen-Frank minimizer. Furthermore, in spherical droplets, the hedgehog configuration becomes unstable \cite{gartland_instability_1999} at low temperatures or large droplet radii for most values of material elastic constants of the liquid crystal.  Mkaddem and Gartland \cite{mkaddem_fine_2000,mkaddem_numerical_1998} carried out some of the first systematic numerical investigations of point defects in nematic liquid crystals. They derived stability conditions for several canonical defect configurations and confirmed these through computations. Their results provided the first systematic numerical exploration of defect stability and transitions, which our work extends by focusing on in detail the bifurcations of split-core and single-core structures.

In this work, we study these defect structures in nematic liquid crystals by minimizing the Landau–de Gennes energy functional. Our focus is on the bifurcation behavior of single-core and split-core defects, and providing a more detailed characterization of the transition between these two configurations. We identify a critical temperature threshold, dependent on domain size, above which the split-core defect ceases to exist. By plotting the critical temperature threshold, we observe asymptotic behavior. We analyze how this configuration transitions into the single-core structure, and also compute and compare the total energies of the single-core and split-core configurations. The size of the defect core is examined in relation to both domain size and temperature, and also the qualitative effects of varying these parameters independently. 



\section{Model}

We consider the two-dimensional domain 
\begin{equation}
    \Omega = \left\{ (x, y) \,\big|\, x^2 + y^2 < \frac{D^2}{4} \right\},
\end{equation}
which is a disc of diameter $D$. This domain can be interpreted as the cross-section of a cylindrical capillary, or more generally, as a slice of any medium with circular cross-section, with the third spatial direction assumed homogeneous.  

The liquid crystal configuration is modeled by the Landau--de Gennes free energy functional, 
\begin{equation}
    E_{LDG} = E_E(\nabla Q) + E_B(Q) 
    = \int_{\Omega} \big( f_E(\nabla Q) + f_B(Q) \big)\, d\Omega,
\end{equation}
whose minimizers determine the equilibrium defect structures. Here $f_E$ denotes the elastic energy density and $f_B$ the bulk energy density, given by
\begin{equation}
\begin{aligned}
    f_E(\nabla Q) &= \frac{L_1}{2}\, Q_{\alpha\beta,\gamma} Q_{\alpha\beta,\gamma}, \\
    f_B(Q) &= \frac{A}{2}\,\mathrm{tr}(Q^2) 
             - \frac{B}{3}\,\mathrm{tr}(Q^3) 
             + \frac{C}{4}\,(\mathrm{tr}(Q^2))^2.
\end{aligned}
\end{equation}
We adopt the one-constant approximation, in which the elastic energy density is governed by a single elastic constant $L_1$. The bulk material parameters $A$, $B$, and $C$ have the following roles: $A$ controls the isotropic--nematic transition, $B$ favors uniaxial alignment, and $C$ penalizes large order parameters. The bulk coefficient $A$ is typically taken to depend linearly on the temperature $T$,  
\begin{equation}
    A \propto T - T_{SC},
\end{equation}
where $T_{SC}$ denotes the supercooling temperature below which the isotropic phase ceases to be locally stable, so that the only minimizers of $f_B$ are uniaxial states. Let $T_{NI}$ be the nematic--isotropic transition temperature and the corresponding bulk coefficient is 
\begin{equation}
    A_{NI} = \frac{B^2}{27C}.
\end{equation}
This motivates the normalized temperature
\begin{equation}
    \theta = \frac{A}{A_{NI}} =\frac{T-T_{SC}}{T_{NI}-T_{SC}}.
\end{equation}
To nondimensionalize the energy functional, we define
\begin{equation}\label{ndconstants}
\begin{aligned}
    \bar{x} &= \frac{x}{D}, \quad \bar{y} = \frac{y}{D}, \quad
    \bar{Q} = \frac{Q}{\alpha}, \quad \alpha = \sqrt{\frac{A_{NI}}{C}}, \quad 
    \bar{\xi}_{NI} = \sqrt{\frac{L_1}{D^2 A_{NI}}}.
\end{aligned}
\end{equation}
With this scaling, the normalized Landau--de Gennes energy functional in two dimensions becomes
\begin{equation}\label{Engeq}
    \frac{E_{LDG}}{A_{NI}\alpha^2 D^2} 
    = \int_{\overline{\Omega}} \left(
    \frac{1}{2}\,\bar{\xi}_{NI}^2 |\nabla_{\bar{\mathbf{x}}} \bar{Q}|^2
    + \frac{\theta}{2}\,\mathrm{tr}(\bar{Q}^2)
    + \frac{1}{4}\,(\mathrm{tr}(\bar{Q}^2))^2
    \right)\, d\bar{\mathbf{x}}.
\end{equation}
In two dimensions, the order tensor $Q$ takes the form
\begin{equation}
Q = \begin{bmatrix}
q_0 & q_1 \\
q_1 & -q_0
\end{bmatrix},
\end{equation}
so that $\mathrm{tr}(Q^3) = 0$ and the cubic term in the energy vanishes. This nondimensionalization is equivalent to that in \cite{gartland_scalings_2018}.  The corresponding Euler-Lagrange equations, in weak form, are
\begin{equation}\label{weakform}
	\left\{
	\begin{array}{l}
    \displaystyle \int_{\Omega} \Big( 2\bar{\xi}_{NI}^2 \nabla q_0 \cdot \nabla v_0
    + (2\theta q_0 + 4(q_0^2+q_1^2)q_0)\, v_0 \Big)\, d\mathbf{x} = 0, \\ \\
    \displaystyle \int_{\Omega} \Big( 2\bar{\xi}_{NI}^2 \nabla q_1 \cdot \nabla v_1
    + (2\theta q_1 + 4(q_0^2+q_1^2)q_1)\, v_1 \Big)\, d\mathbf{x} = 0, 
	\end{array}
	\right.
\end{equation}
where we have dropped the overbars for simplicity, and $v_0$ and $v_1$ are admissible test functions corresponding to the unknown functions $q_0$ and $q_1$, respectively.

\section{Numerics}
The numerical simulations in this study are based on the material properties of the liquid crystal 5CB, with relevant constants listed in Table~\ref{t1}.  

\begin{table}[!h]
\centering
\caption{Material constants for liquid crystal 5CB}\label{t1}
\begin{tabular}{|l|l|l|l|}
\hline
\textbf{Parameter} & \textbf{Value} & \textbf{Parameter} & \textbf{Value} \\ \hline
$B$ & $-2.12 \times 10^6\ $J/m\(^3\) \cite{ravnikEntangledNematicColloidal2007} & $L_1$ & $4.00 \times 10^{-11}$ J/m \cite{ravnikEntangledNematicColloidal2007} \\ \hline
$C$ & $1.73 \times 10^6\ $J/m\(^3\) \cite{ravnikEntangledNematicColloidal2007} & $T_{SC},\ T_{NI}$ & $22.5,\ 35.4\ $°C \cite{stewartStaticDynamicContinuum2019a} \\ \hline
\end{tabular}
\end{table}

Throughout this work, we adopt the calculated value $A_{NI}\alpha^2  \approx 5352\ \mathrm{J/m^3}$ for the nondimensionalization from Eq. \ref{ndconstants}. Under this scaling, the normalized temperature $\theta$ is given by
\begin{equation}\label{theta_T_relation}
\theta = \frac{T-T_{SC}}{T_{NI}-T_{SC}} \approx 0.0775 \,(T-22.5),
\end{equation}
where $T$ is expressed in °C.  

The weak form equations \eqref{weakform} are solved on a domain discretized using a mesh generated with the \texttt{Gmsh} package, and computations are performed with the finite element software Firedrake. The resulting nonlinear systems are solved using Newton’s method, which requires a suitable initial guess to ensure convergence. For initial guess, we begin with constant initial guess $(q_0,q_1)=( \frac{1}{\sqrt{2}},\frac{1}{\sqrt{2}})$, and also make use of split core and planar radial initial guesses throughout this work. The planar radial configuration \cite{dassbach_computational_2017} is an explicit solution to the Euler-Lagrange equations. Any split-core initial guess we use is obtained via saving a split core resulting from either a constant initial guess or previous split core result, and using this split-core configuration as the new initial guess. The computed fields $q_0$ and $q_1$ provide detailed information on the molecular orientation, defect structures, and the spatial distribution of uniaxial and biaxial ordering throughout the domain.

\subsection{Convergence Analysis}
For our discretization of the 2-D disc, we begin with approximately 3,000 triangular elements. The mesh is progressively refined, quadrupling the number of triangles with each successive refinement. We monitor energy convergence throughout these steps to ensure accuracy by comparing successive energy values. For this we set the fixed domain diameter to $D = 1 \times 10^{-7}$\ m. The parameters $\theta$ and the initial guesses are selected to sample a broad range of physical configurations in the following, and using different initial guesses allows us to obtain the different defect configurations, due to the Newton method's dependence on initial guess.

\begin{figure}[H]
		\centering
		\includegraphics[width=6.0cm, height=4.9cm]{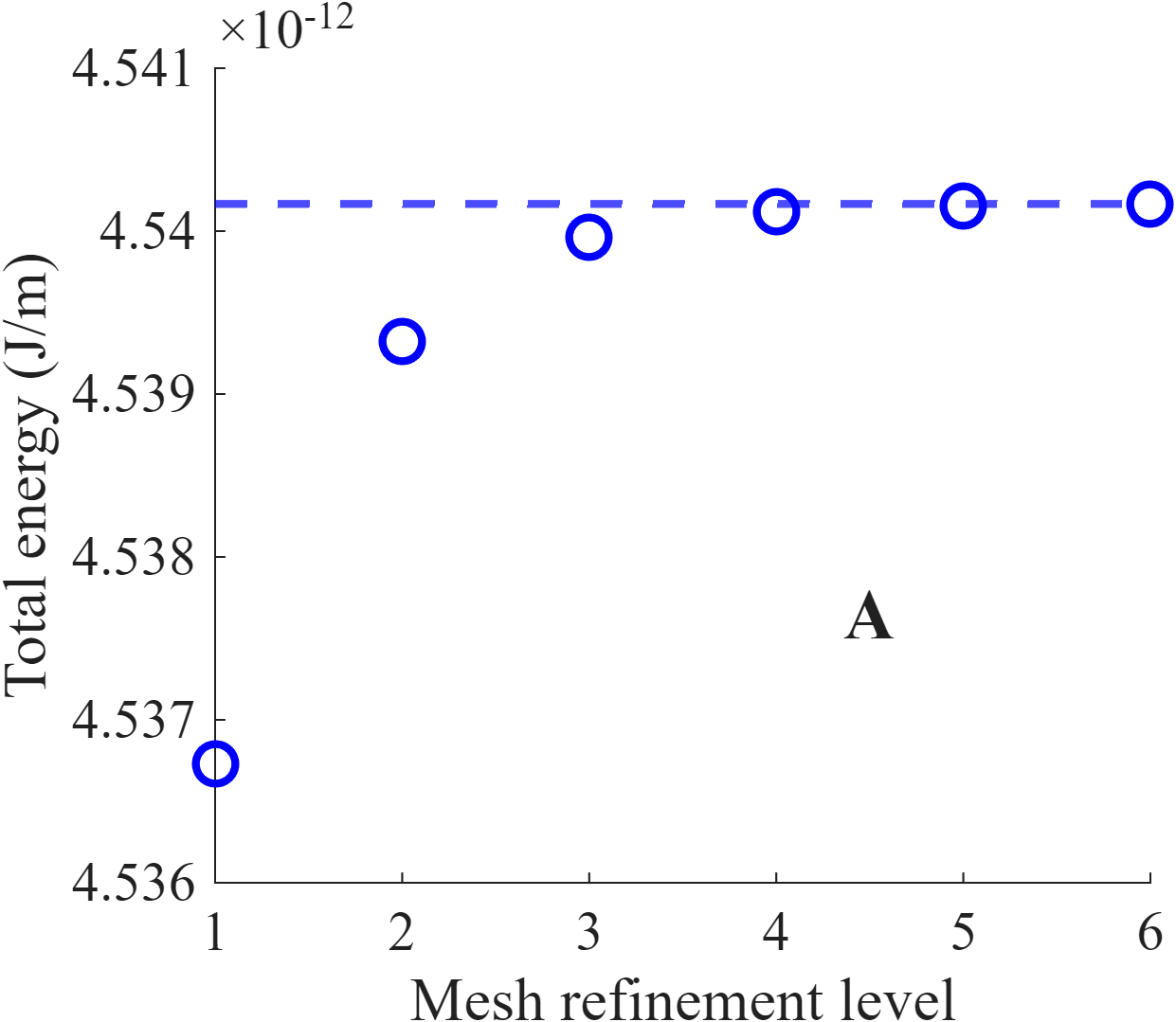}
        \includegraphics[width=6.0cm, height=4.9cm]{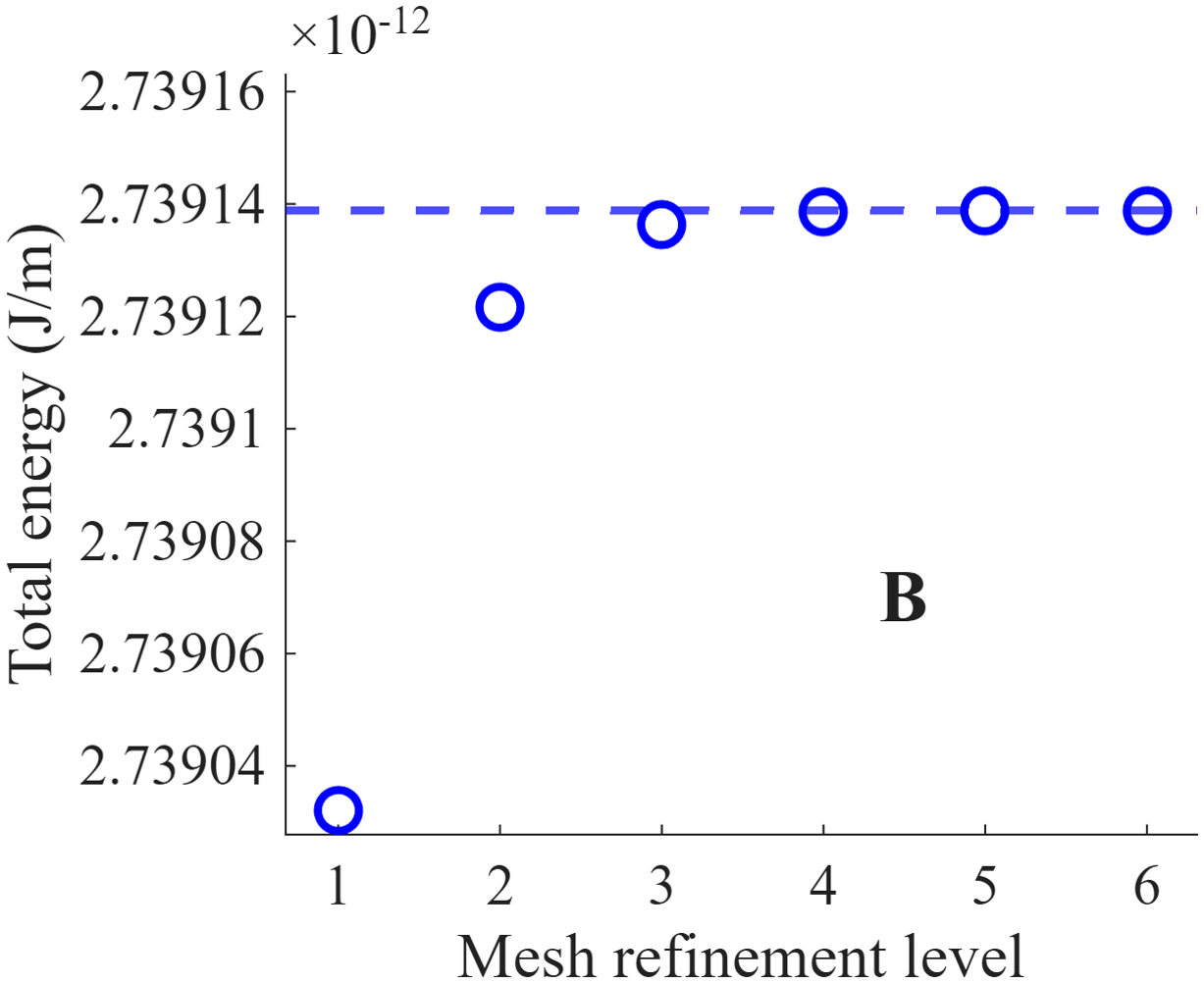}
\includegraphics[width=6.0cm, height=4.9cm]{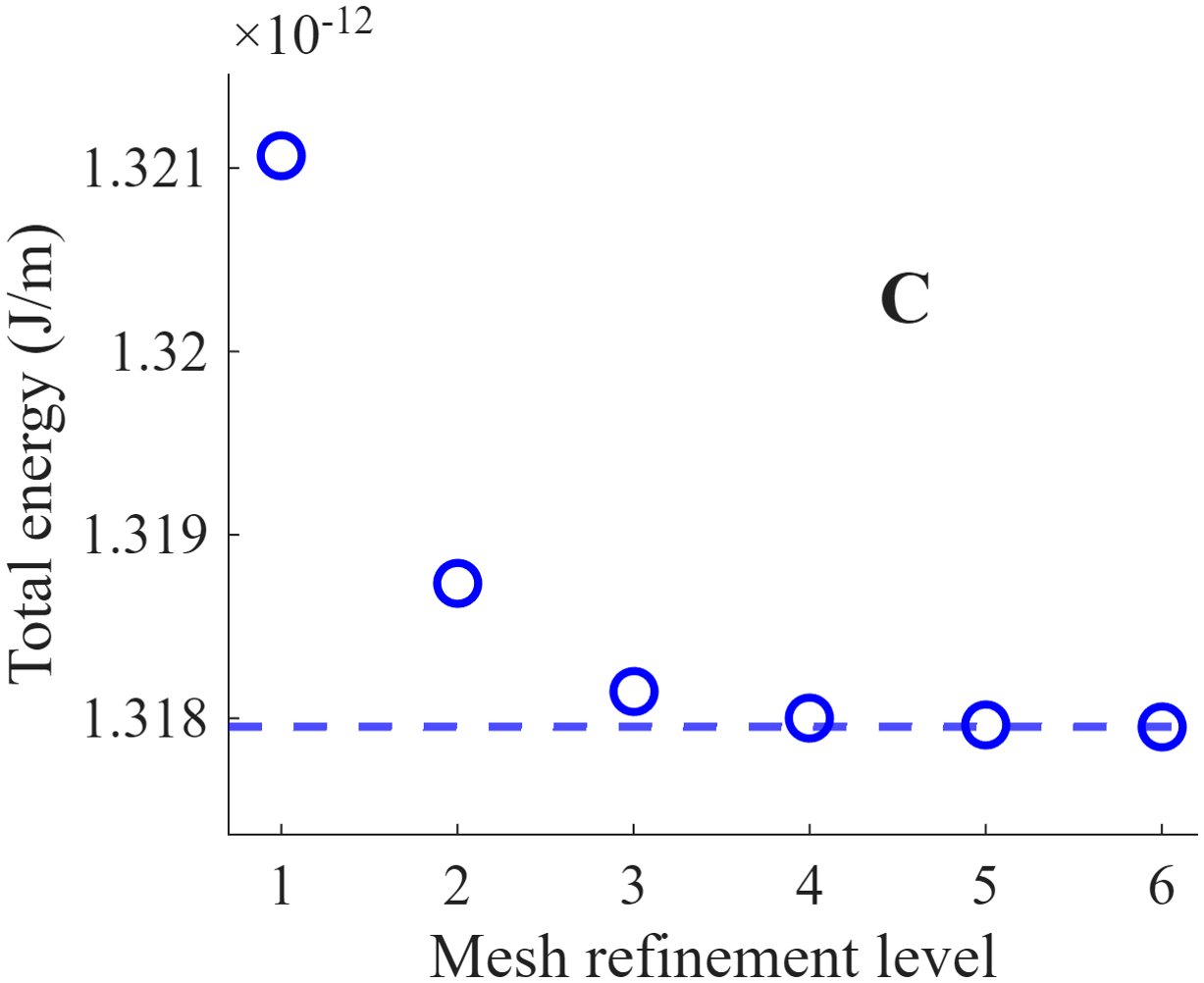}
        \includegraphics[width=6.0cm, height=4.9cm]{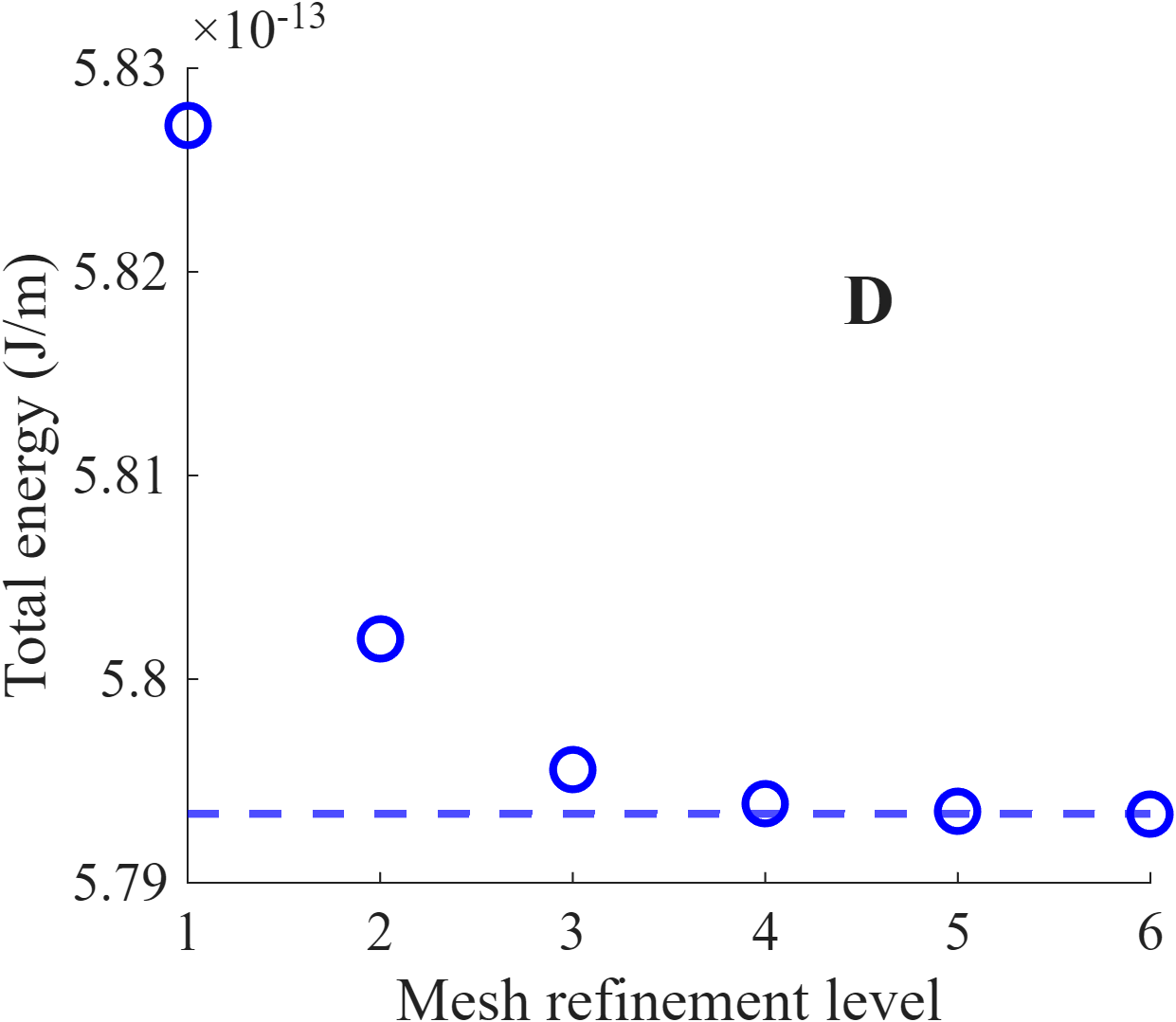}\caption{Total energy for domain diameter $D = 1 \times 10^{-7}\ $m. \textbf{A}: Single core for $\theta = -0.8 ,$ (12.18 °C); \quad \textbf{B}: Split core for $\theta = -1.2 ,$ (7.016 °C); \textbf{C}: Single core for $\theta = -1.5, $ (3.145 °C);\quad \textbf{D}: Split core for $\theta = -1.5, $ (3.145 °C). The total energy $E_{LDG}$ is calculated using Eq. \eqref{Engeq}. Planar radial (single core) initial guess is used for A and C, and split core initial guess is used for B and D.}\label{convgraph}
	\end{figure}

Successive mesh refinements demonstrate consistent convergence of the computed total energy. Connecting the total energy values obtained from consecutive refinements produces line segments that become more horizontal with each successive mesh refinement, which indicates diminishing differences between successive energy values. This behavior confirms that the computed total energy converges towards the true value as mesh resolution increases, as demonstrated by Figure \ref{convgraph}.

\subsection{Total energy comparisons}
We conduct a broad set of simulations across different $D$ values, $D$ being at typical scales for which the aforementioned liquid crystal defect types can be observed relative to the domain size. For each choice of $D$, we compute the total energy of both the single-core and split-core configurations over a temperature range that supports the nematic liquid crystal phase. Planar radial initial guesses are expected to generate single core figures, and split core initial guesses are expected to generate split core figures. The process we use for total energy comparisons is the following. We fix domain size $D$ and begin with a higher temperature inside the range at which the material can be found in liquid crystal form, such as $0 \leq \theta \leq 1 $. We employ the constant initial guess and compute defect configurations and their corresponding energies for decreasing temperatures until the ``order parameter" violates physical constraints. At some temperature during this process, we observe that the result changes from single-core to split-core configuration. Then, we begin from below this transition point and increase temperature, using a split core result as the initial guess, until the split core configuration can no longer be obtained. Thirdly, we use planar radial initial guess to obtain remaining total energies for the single-core configuration below the transition point. The mesh we use for these total energy calculations is comprised of about three million similarly-sized triangular elements.

For a typical domain size such as $D \approx 1.5 \times 10^{-6}$ m, for example, both the single core and split core defects are observed up to $\approx$ 22 °C. Across different domain sizes, including $D = \{2 \times 10^{-8},\ 1 \times 10^{-7},\ 5 \times 10^{-7},\ 2.5 \times 10^{-6}\}\ $m, the qualitative behavior is very similar. However, for smaller domains, such as when the domain size is $D = 2 \times 10^{-8}\ $m, the coexistence temperature range falls below the regime in which the material remains in the nematic liquid crystal phase. It is possible that in smaller domains like these, the diameter is not sufficiently large to support the existence of both of these defect configurations. For larger domains, both types of defect consistently coexist at temperatures near the isotropic lower limit, $T_{SC}$. Moreover, increasing the domain diameter $D$ results in defect cores that occupy a smaller fraction of the domain.

In what follows, we present the detailed results for $D=1 \times 10^{-7}$ m.
\begin{figure}[!h]
		\centering
		\includegraphics[width=12.0cm, height=6.72cm]{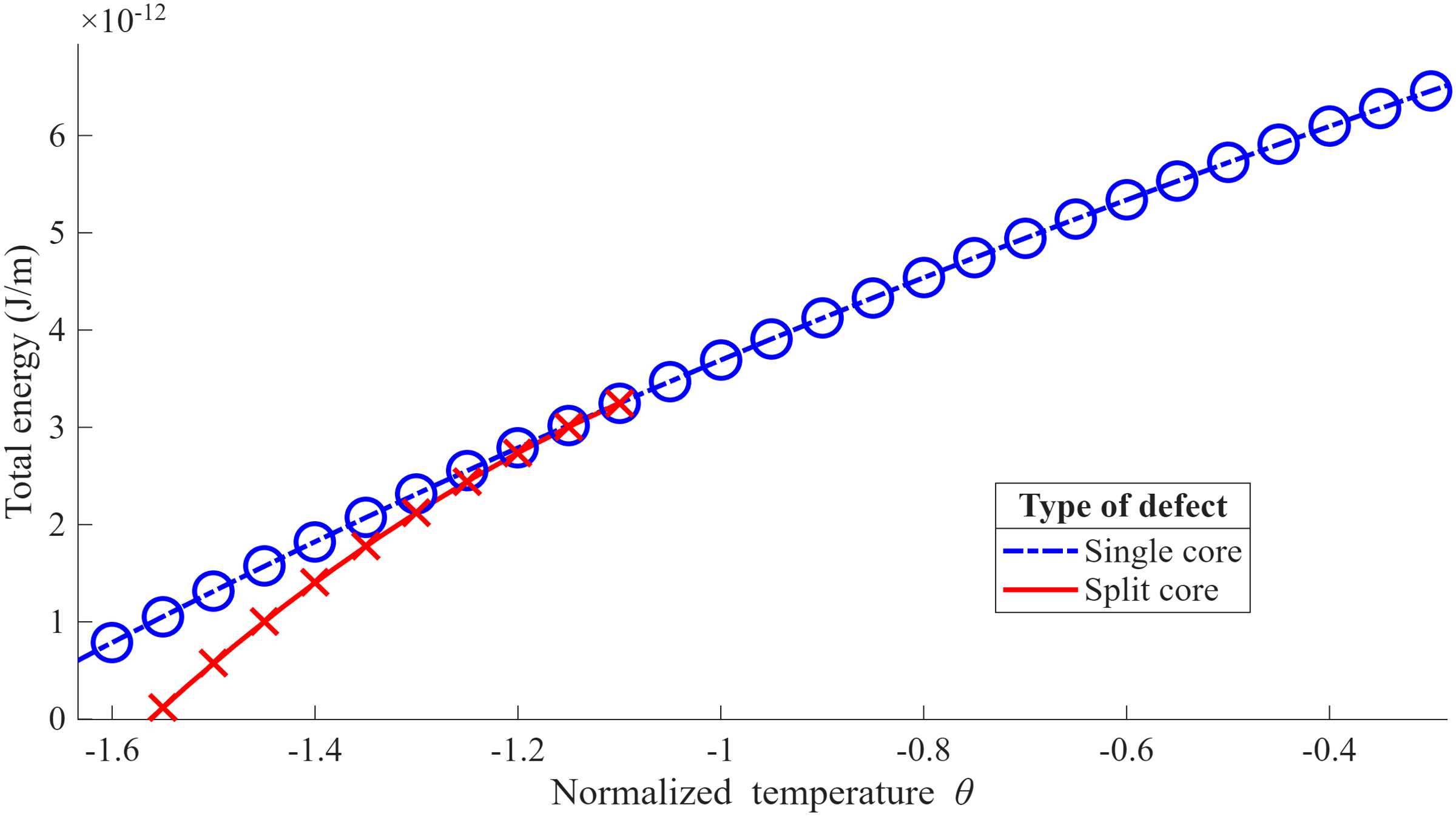}
		\label{fig:res7}\caption{Energy calculated over 2-D disc, $D=1 \times 10^{-7}$ m. Single core and split core results. For this domain size, the split core defect was not observed for $\theta > -1.10\approx 8.306$ °C. The total energy $E_{LDG}$ is calculated using Eq. \eqref{Engeq}.}
	\end{figure}
The single-core defect is found to exist across the entire simulated range of $\theta$. In contrast, the split-core defect, even for domain diameters $D$ large enough to permit its stability, is observed only below a critical temperature that depends on the selected value of $D$. According to all our efforts including trying various initial guesses, we did not see the split core configuration for any temperatures higher than this critical temperature. Below this temperature threshold, both configurations are possible, and the resulting defect type depends on the given initial guess. At the upper temperature limit of where the split-core remains viable, the total energies of the single-core and split-core configurations are nearly indistinguishable. As the temperature decreases further, however, the energy of the split-core configuration decreases much more rapidly than that of the single-core configuration, although each configuration exhibits the expected monotonic decrease with decreasing temperature.

\subsection{Effect of increasing $\theta$}
For a uniaxial liquid crystal in two dimensions, the scalar order parameter is defined as \begin{equation}s_h := 2\lambda_h,\end{equation} where $\lambda_h$ denotes the dominant eigenvalue of the $Q$-tensor. Any point with $s_h = 1$ corresponds to perfect uniaxial alignment at that location. Values of $s_h$ close to zero typically indicate the presence of a defect core or a disordered region. Therefore the defect configuration can be identified by computing and visualizing $s_h$ over the domain. We set $D = 1 \times 10^{-7}$\ m, discretized using a mesh of approximately three million triangular elements, and increase the temperature beginning from within the range where the split-core defect exists. This approach enables us to track the evolution of the split-core defect as the system transitions toward the single-core configuration. \begin{figure}[!h]
		\centering
		\includegraphics[width=6.04cm, height=3.4cm]{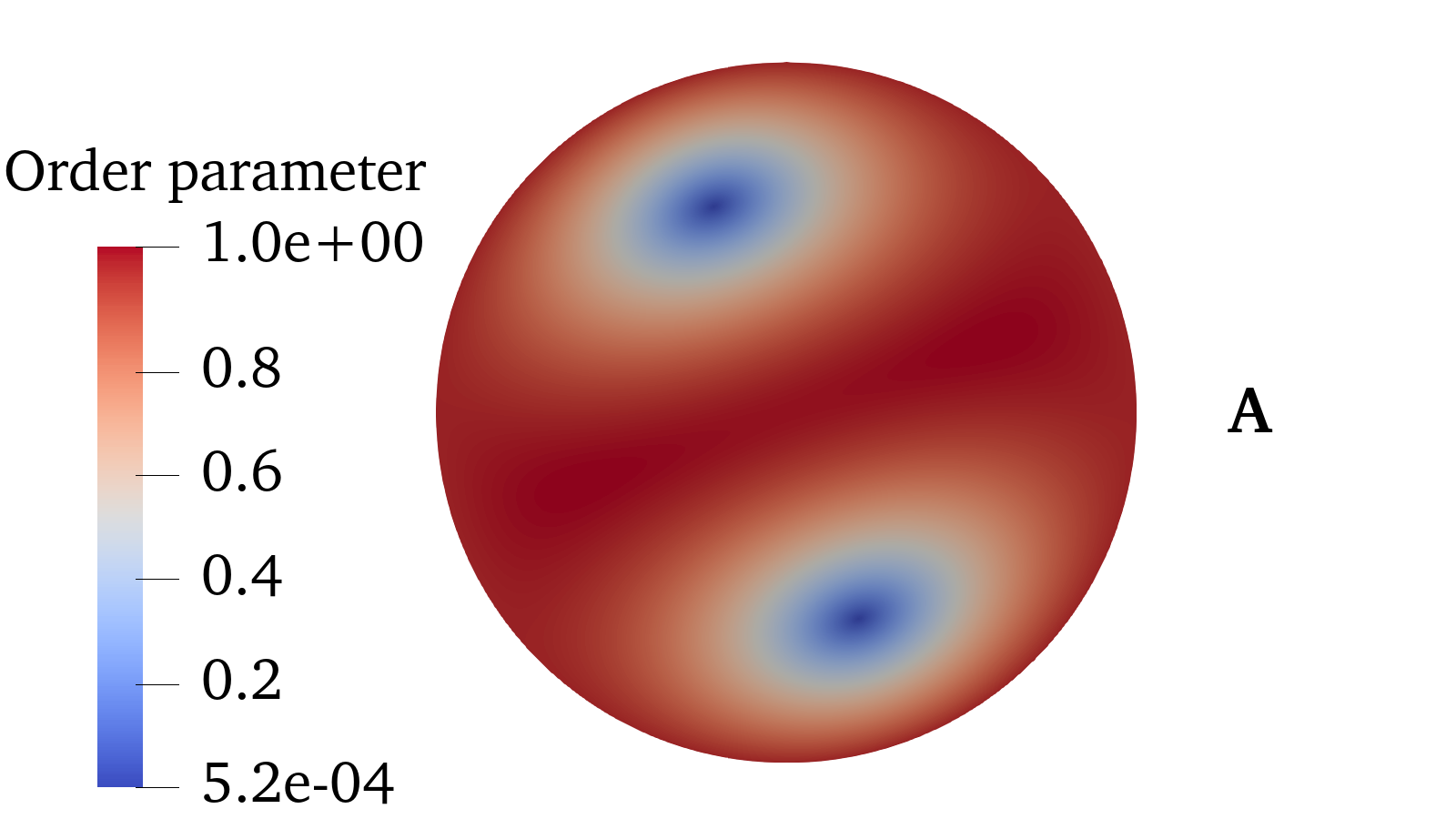}
        \includegraphics[width=6.04cm, height=3.4cm]{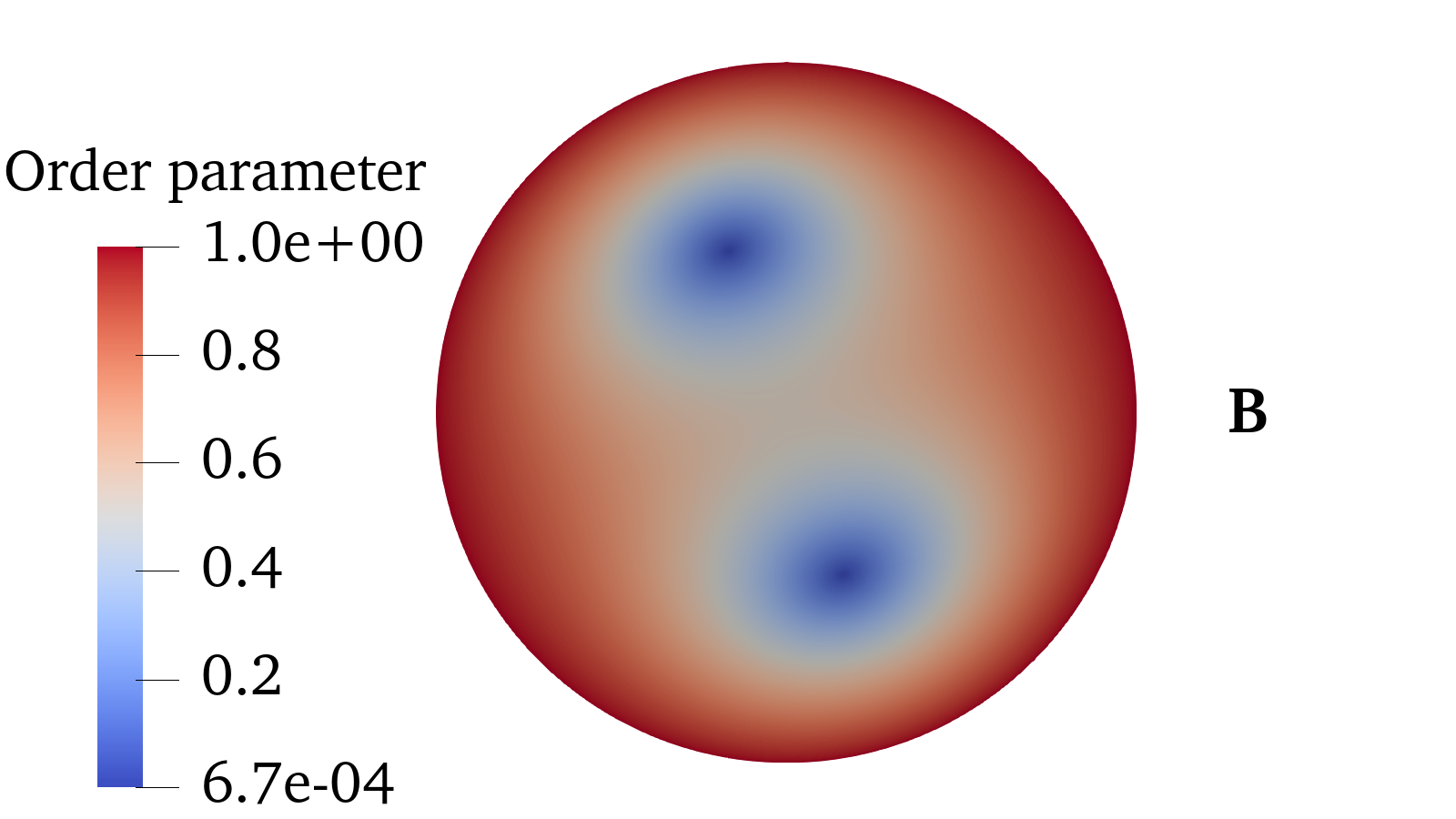}
        \includegraphics[width=6.04cm, height=3.4cm]{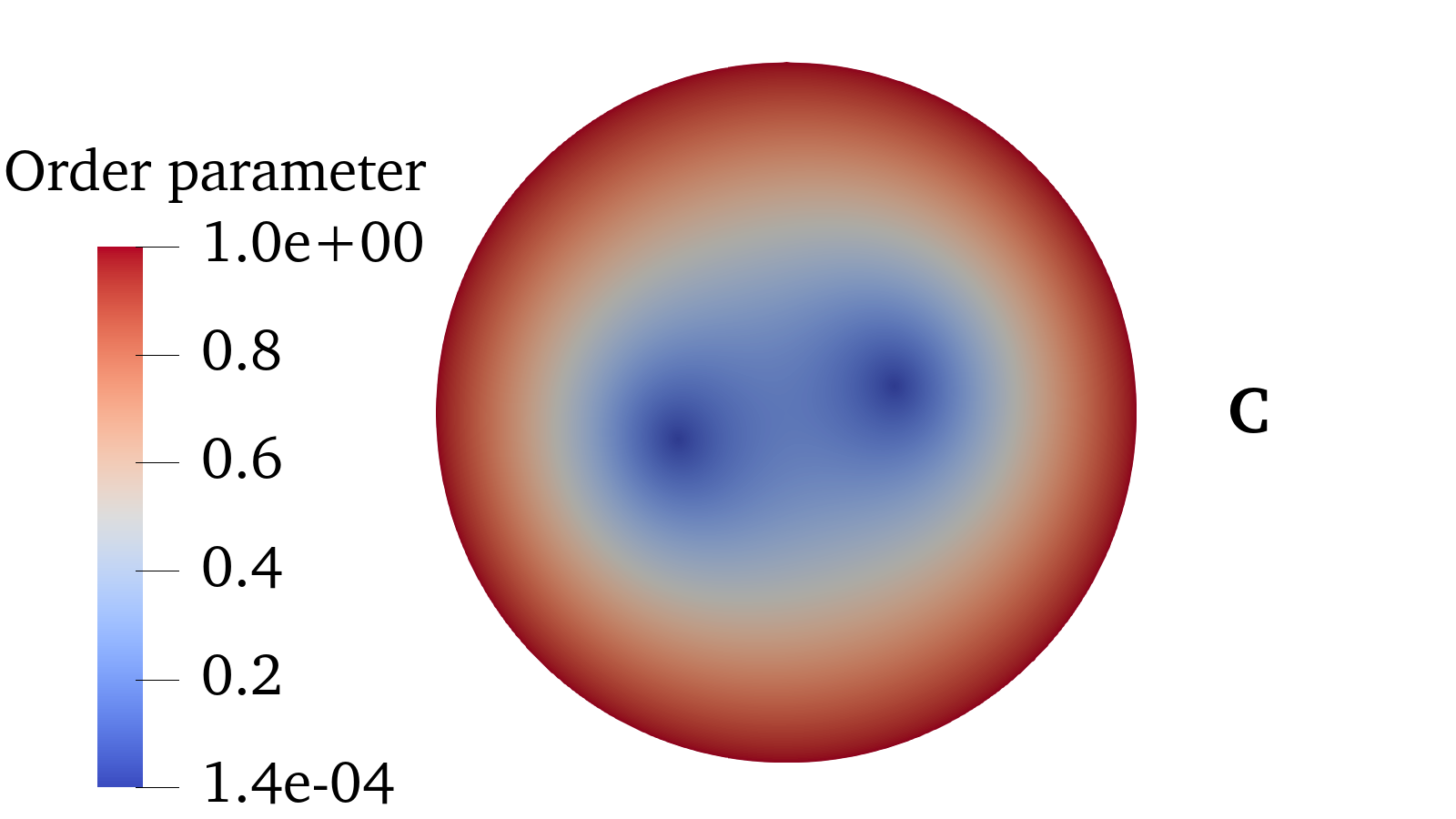}
        \includegraphics[width=6.04cm, height=3.4cm]{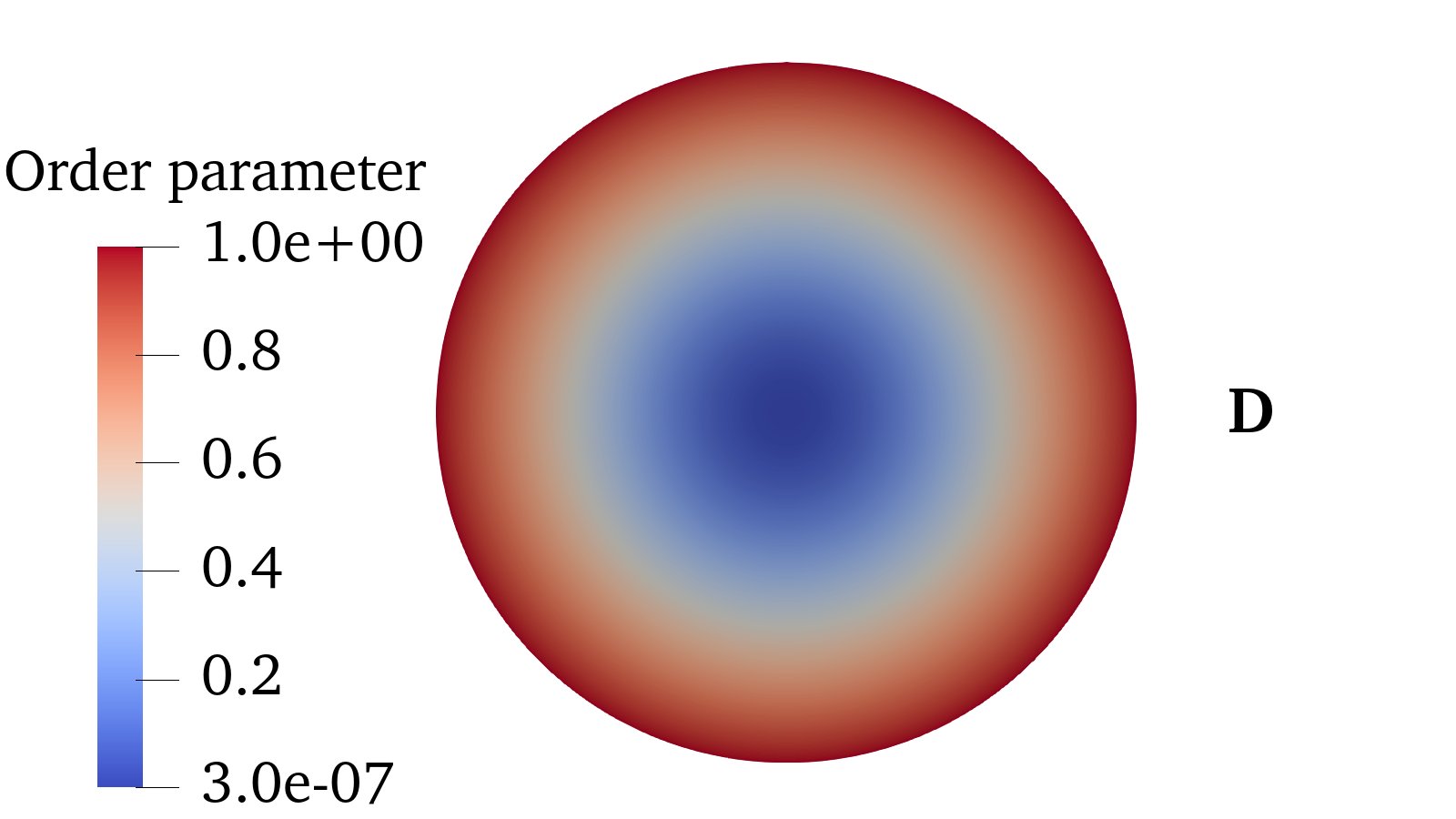}
		\label{fig:t2o}\caption{Plots of scalar order parameter $s_h = 2 \lambda_h$ for different temperatures with domain diameter $1 \times 10^{-7}$\ m. \textbf{A}: $\theta = -1.4 $ (4.435 \text{°C}), \ \textbf{B}: $\theta = -1.17 $ (7.403 \text{°C}),\ \textbf{C}: $\theta = -1.1 $ (8.306 \text{°C}), \ \textbf{D}: $\theta = -1.05 $ (8.952 \text{°C}). Split core initial guess is used.}
	\end{figure} 
As the temperature increases, the two cores of the split-core configuration gradually approach one another until they merge into a single-core defect located at the center of the domain. Once this transition occurs, the resulting single-core defect exhibits a noticeably larger core size compared to either of the two constituent cores of the initial split configuration.

To further examine the influence of domain size, we compute and visualize the scalar order parameter $s_h$ for a significantly larger domain, $D = 1.52 \times 10^{-6}$\ m. The simulations are again initialized within the temperature range where the split-core defect can be found. We again use a three million element mesh.
\begin{figure}[!h]
		\centering
		\includegraphics[width=6.0cm, height=3.4cm]{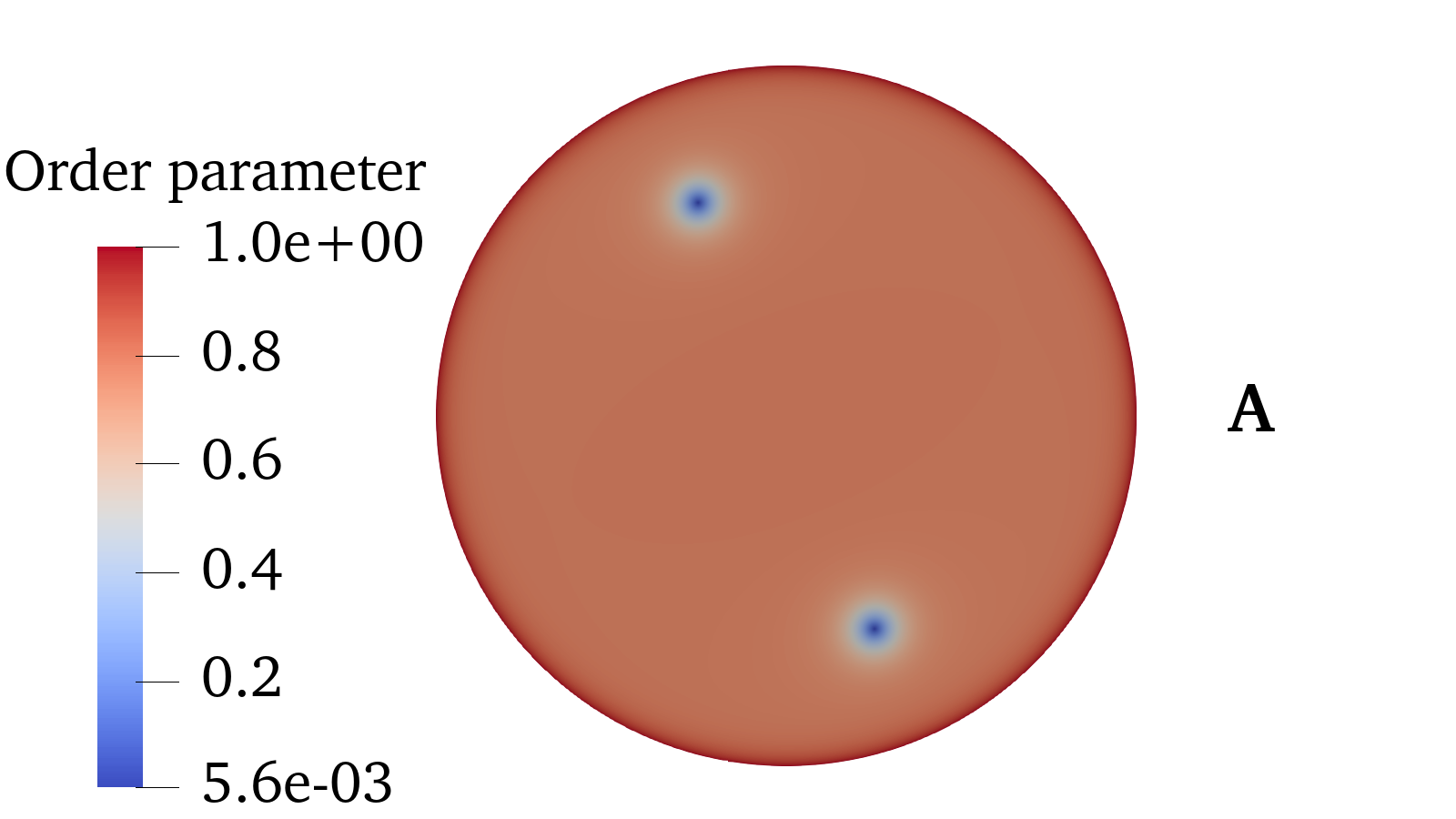}
        \includegraphics[width=6.0cm, height=3.4cm]{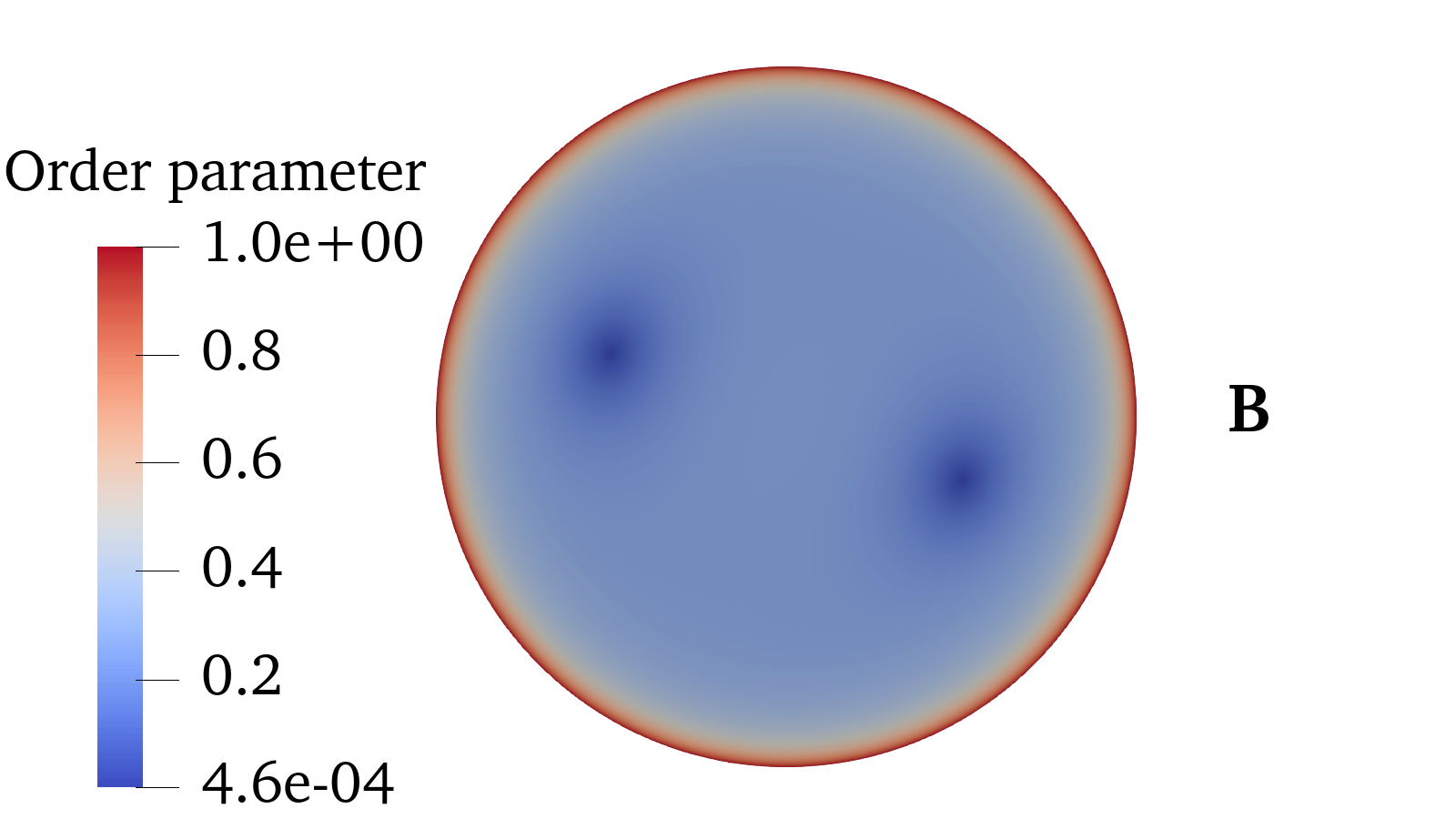}
        \includegraphics[width=6.0cm, height=3.4cm]{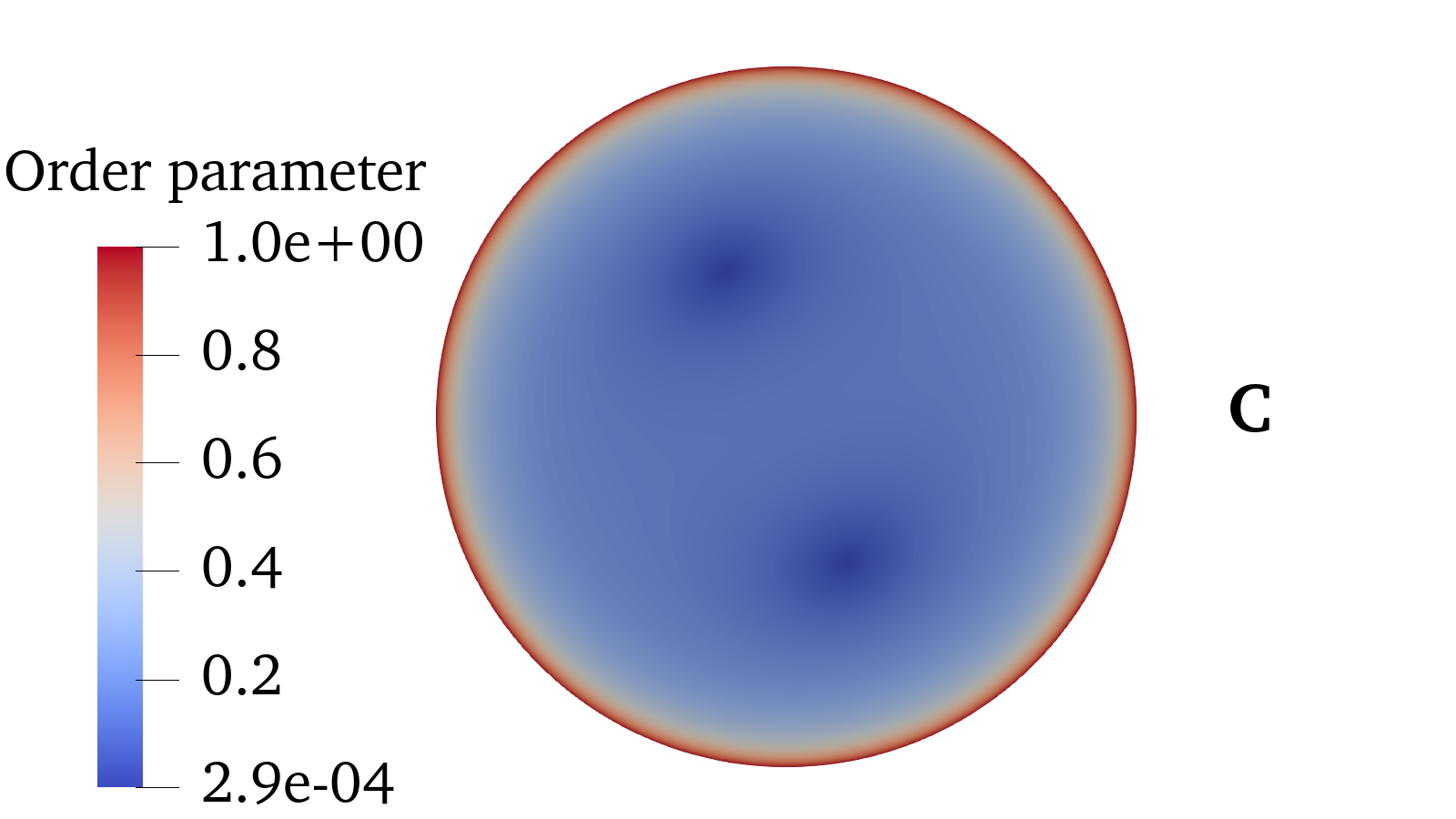}
        \includegraphics[width=6.0cm, height=3.4cm]{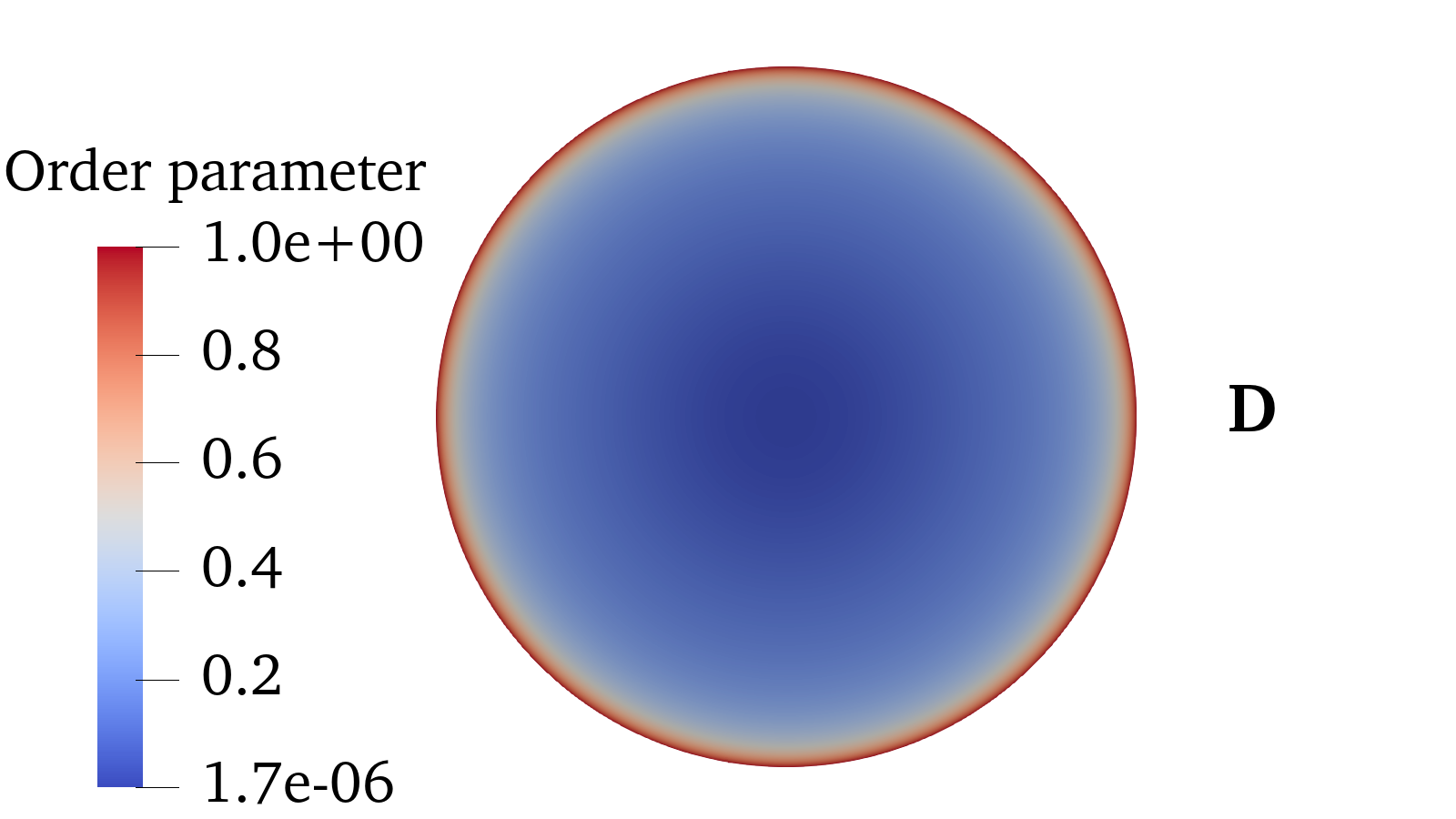}
		\label{fig:t2o2}\caption{Plots of scalar order parameter $s_h = 2 \lambda_h$ for different temperatures with $D = 1.52 \times 10^{-6}$\ m. \textbf{A}:  $\theta = -0.30 $ (18.63 \text{°C}), \ \textbf{B}:  $\theta = -0.04 $ (21.98 \text{°C}), \ \textbf{C}: $\theta = -0.02 $ (22.24 \text{°C}), \ \textbf{D}: $\theta = -0.01 $ (22.37 \text{°C}). Split core initial guess is used.}
	\end{figure}
With the domain diameter being significantly larger, the two defects comprising the split-core configuration are comparatively much smaller relative to the overall size of the domain. However, as the temperature increases toward the upper limit of the split-core existence range, the individual cores expand noticeably in size. This behavior contrasts with that observed in smaller domains, where the two defects of the split-core configuration exhibit little change in size until merging into the single-core defect. Moreover, in larger domains, once the transition to the single-core configuration occurs, the resulting core is disproportionately large, even relative to the already enlarged domain size.

\subsection{Effects of domain size on split core threshold}
Using meshes consisting of approximately 800,000 elements, we determine the highest temperature at which the split-core solution persists for a range of domain sizes. We denote this temperature as $\theta_{tr},$ the ``split core threshold". This threshold is computed using a bisection procedure with multiple subdivisions at each refinement step. After each step, we identify the maximum value of $\theta$ at which the split-core configuration is present by examining the corresponding plots of $s_h$, all with split-core initial guess. The process is then repeated on the temperature interval bounded by this $\theta$ and the next higher $\theta$. This iterative refinement continues until the maximum relative error satisfies $\theta_{err} < 5\%$, for each choice of $D$. The resulting values of $\theta_{tr}$ as a function of $D$ provide a quantitative measure of how domain size influences the stability of the split-core configuration.\begin{figure}[!h]
		\centering
        \includegraphics[width=12.0cm, height=6.72cm]{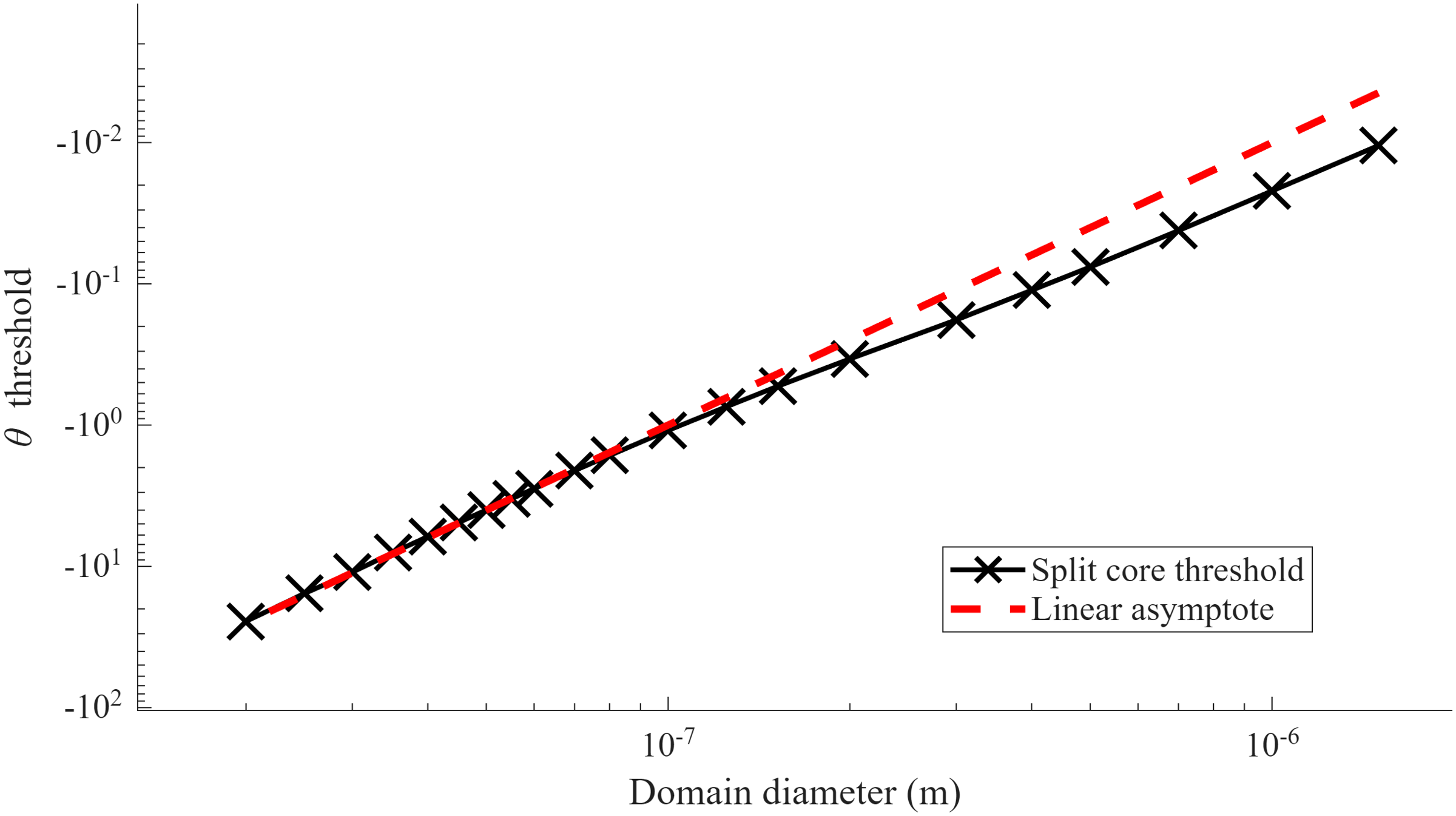}
		\label{fig:thrvssize}\caption{ Interpolation of split core threshold for different domain diameters $D$ and linear asymptote calculated via Eq. \eqref{st2}.}
	\end{figure} 

When plotted on logarithmic axes, the relationship between the domain diameter $D$ and the split-core threshold $\theta_{\mathrm{tr}}$ is observed to be approximately linear. We perform a linear fitting using this data and this yields
\begin{equation}\label{st2}
  \log(-\theta_{\mathrm{tr}}) \approx -2 \log(10^7 D), \ \ \ \theta_{\mathrm{tr}} \approx -(10^7D)^{-2}.
\end{equation}
This approximation is valid for small domain sizes, though slight deviations are observed for larger domains.

\subsection{Defect core size}
To quantify the size of defect cores, we perform computations on meshes consisting of approximately $200{,}000$ triangular elements. The defect region is identified by applying a Heaviside function to the scalar order parameter $s_h$, with cutoff threshold $s_h < 0.1$, defined as
\begin{equation}\label{Heav1}
  \mathcal{H}_{0.1}(\mathbf{x}) :=
  \begin{cases}
    1, & s_h(\mathbf{x}) < 0.1, \\
    0, & s_h(\mathbf{x}) \geq 0.1.
  \end{cases}
\end{equation}
The normalized defect area $S_{\mathrm{n}}$ is then computed as
\begin{equation}
  S_{\mathrm{n}} = \int_{\overline{\Omega}} \mathcal{H}_{0.1}(\bar{\mathbf{x}})\,d\bar{\mathbf{x}},
\end{equation}
and the corresponding absolute defect area $S$ is obtained by scaling with the physical domain size. Since only a discrete version of $\mathcal{H}_{0.1}$ can be implemented, an error is introduced due to the finite mesh resolution. Define
\[
  \Omega_s(\mathbf{x},\mathbf{\Omega}) := \{\, \mathbf{x} \in \mathbf{\Omega} : s_h(\mathbf{x}) < 0.1 \,\}.
\]
If the boundary $\partial \Omega_s$ does not coincide with the mesh element boundaries, the discrete $\mathcal{H}_{0.1}$ differs from its continuous analogue. The resulting discretization error is of order $m/2$, where $m$ denotes the size of an individual mesh element. For a single-core defect, $\Omega_s$ is approximately circular; its radius can be estimated as
\[
  r \approx \sqrt{\tfrac{S_{\mathrm{n}}}{\pi}} \;\; \pm \tfrac{m}{2}.
\]
Consequently, multiplying $\pi r^2$ by $D^{2}$ gives bounds for the absolute defect area:
\begin{equation}\label{abscorearea}
  D^2\pi \left( \sqrt{\tfrac{S_{\mathrm{n}}}{\pi}} \;\; - \tfrac{m}{2} \right)^{2} \leq S \leq D^{2}\,\pi \left( \sqrt{\tfrac{S_{\mathrm{n}}}{\pi}} \;\; + \tfrac{m}{2} \right)^{2},
\end{equation} We examine defect core size first for varying $\theta$ and then for varying $D$.

\begin{figure}[!h]
		\centering
        \includegraphics[width=12.0cm, height=6.72cm]{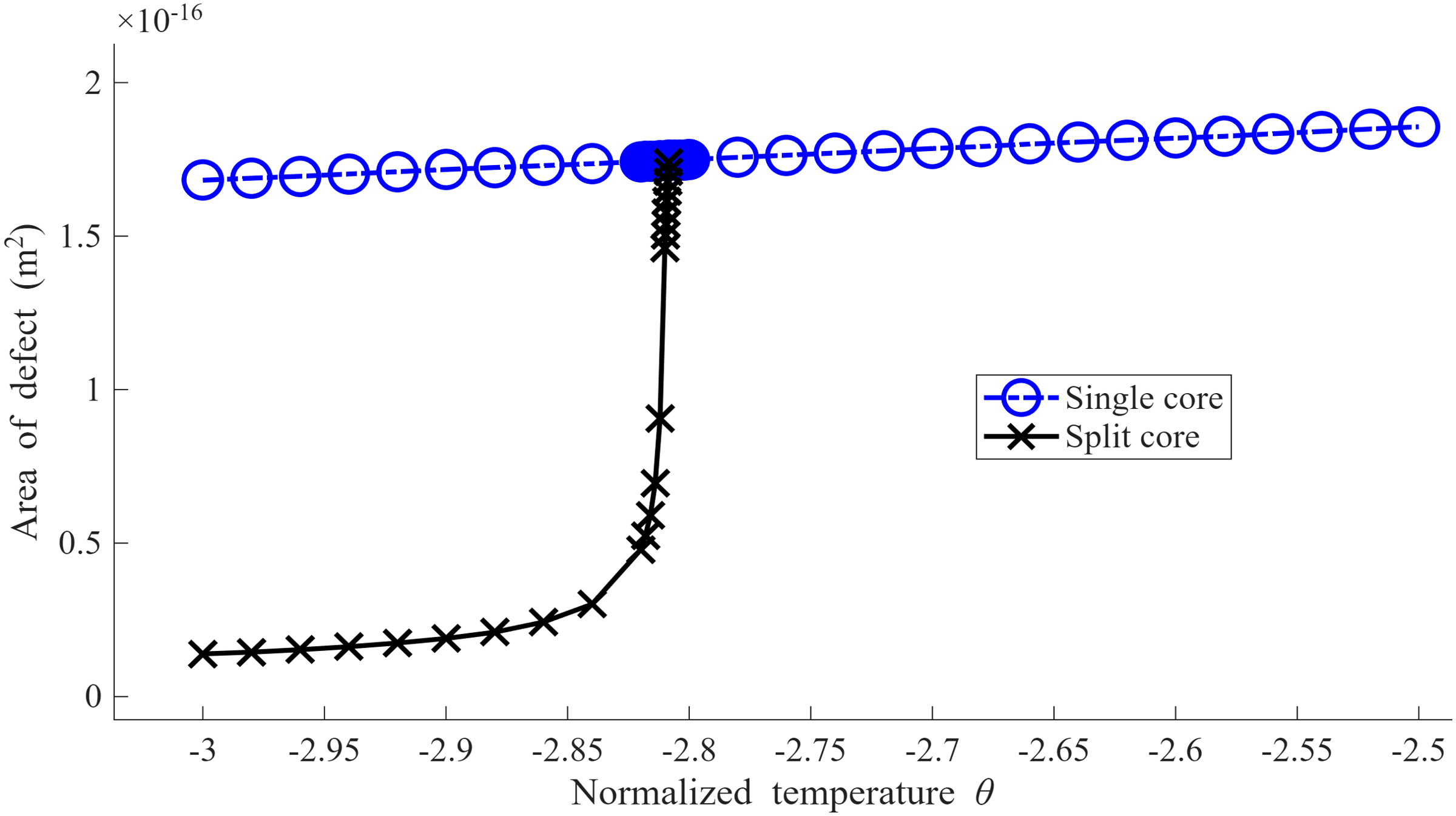}
		\caption{Single core and split core defect size, $D = 6 \times 10^{-8} $ m, varying $\theta$. Single-core configurations are obtained using planar radial initial guess. Split-core configurations are obtained using split-core initial guess. We use Eq. \eqref{abscorearea} to compute the defect area \textit{S}.}\label{fig:TvarRfixed}
	\end{figure}
In Fig. \ref{fig:TvarRfixed}, we observe that for a single-core defect located at the center, the defect area increases approximately linearly as the temperature is raised at a constant rate. In the case of a split-core defect, starting at temperatures below the split-core threshold, the defect area initially increases with temperature at a rate comparable to that of the single-core configuration. However, as the temperature continues to rise, the area of the split-core defect exhibits accelerated growth, producing a parabola-like curve as it approaches the split-core threshold from below. Near this threshold, the rate of increase shows no sign of slowing until the split-core defect coalesces into the single-core configuration. Similar qualitative behavior is observed for other fixed values of $D$, including in the vicinity of their respective split-core thresholds.

\begin{figure}[!h]
		\centering
        \includegraphics[width=12.0cm, height=6.72cm]{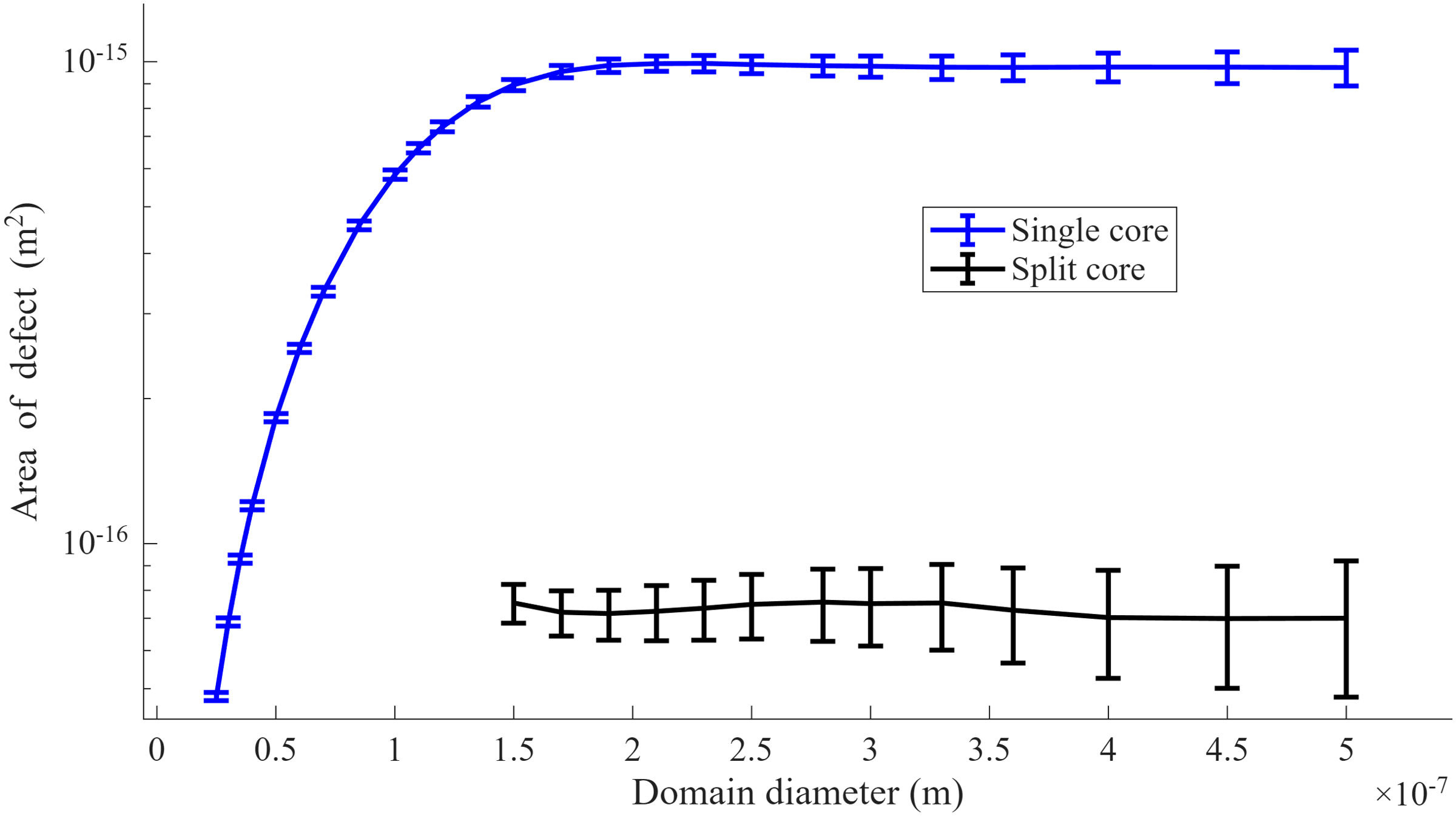}
		\caption{Single core and split core defect size, $\theta = -0.75$\ (12.82 °C), for varying domain diameter. Single-core configurations are obtained using planar radial initial guess. Split-core configurations are obtained using split-core initial guess. Defect area is computed using Eq. \eqref{abscorearea}.}\label{fig:TfixedRvar}
	\end{figure}
We now fix $\theta$ and vary the domain size. Although numerical errors increase slightly with larger domain diameters, several conclusions can still be drawn from the resulting Fig. \ref{fig:TfixedRvar}. For the single-core configuration, increasing the domain diameter $D$ leads to a corresponding increase in defect area until a maximum size is reached, beyond which the core size does not change. In contrast, for the split-core configuration, maintaining a fixed temperature while increasing $D$ has minimal effect on the defect core size. These observations indicate that the single-core defect possesses an intrinsic maximum size once it is no longer constrained by a small domain or by proximity to the domain boundary under radial anchoring conditions. The split-core defect exhibits an intrinsic core size at this fixed temperature which appears to be largely independent of domain size.

\section{Discussion}
The numerical results presented in this work establish a clearer picture of the stability and transition of defect structures in nematic liquid crystals under geometric confinement. At low temperatures, both split-core and single-core configurations satisfy the Euler-Lagrange equations, with the split-core configuration exhibiting lower energy. Consequently, within the temperature regime where either defect can exist, the split-core represents the energetically favorable state. As the temperature increases, the two cores comprising the split defect move progressively closer together and eventually merge to become a single-core defect located at the domain center. At sufficiently high temperatures, only the single-core defect can be observed.  

A critical finding of this study is the estimation of the temperature threshold $\theta_{\mathrm{tr}}(D)$, a function of domain size, above which the split-core defect ceases to exist. Numerical computations demonstrate that this threshold is accurately captured by the scaling law in Eqs.~\eqref{st2}, where $D$ denotes the domain diameter in meters. This scaling relation reflects that $\theta_{tr}$ decreases monotonically with decreasing $D$, indicating that confinement and proximity of the boundary under these conditions strongly suppress the stability of split-core defects. From a physical standpoint, this highlights the sensitivity of defect stability to domain size. Specifically, smaller systems require lower temperatures to stabilize split-core structures relative to single-core structures.

In addition to this threshold function, the computations reveal important information about defect core size. As the split-core configuration approaches its upper existence limit in temperature, the defect core size increases significantly. During the transition from the split-core to the single-core configuration, the merging of the two cores is accompanied by a marked growth in core size, consistent with the numerical measurements of defect area in Fig. \ref{fig:TvarRfixed}. 
For fixed $\theta$, increasing the domain diameter enlarges the core of the single-core defect until a maximum size is reached, while in contrast the split-core defect exhibits little dependence on domain size in this respect.  

These findings complement and extend prior analytical and numerical studies of defect stability in nematic systems, particularly by providing a quantitative description of the threshold law and a detailed characterization of defect core size near the transition. At the same time, several limitations of the present framework must be acknowledged. The analysis has been carried out with the one-constant approximation, and the choice of boundary conditions has been restricted to homeotropic only. Moreover, the estimation of defect core size relies on a discretized Heaviside function, introducing errors proportional to the mesh resolution.  

Future work could naturally extend this study in several directions. Including more of the elastic constants rather than using the one-constant approximation would provide a more complete description of material-dependent effects. Exploring alternative boundary conditions or domain geometries would allow for broader applicability. Incorporating dynamical effects into the model could yield more accurate core structure determination, at the cost of greater complexity of the model. 
Finally, a direct comparison between the present numerical results and experimental measurements of confined nematic systems would be of significant interest.  

Taken together, these results highlight how confinement effects and temperature jointly determine the scaling laws and stability regimes of nematic defects. The numerical analysis provides both a quantitative scaling law for the stability threshold and qualitative insight into the mechanisms underlying defect transitions, thereby contributing to a deeper understanding of confinement effects in liquid crystal physics.

\bibliographystyle{siam}
\bibliography{refer}

\end{document}